\documentclass[fleqn,usenatbib]{mn2e}
\usepackage{mathrsfs}
\usepackage{tipa}
\usepackage{amsfonts}
\usepackage{bbm}
\usepackage{amsmath}
\usepackage{amssymb}
\usepackage{graphicx}

\def\gsim{\;\lower4pt\hbox{${\buildrel\displaystyle >\over\sim}$}\;}
\def\lsim{\;\lower4pt\hbox{${\buildrel\displaystyle <\over\sim}$}\;}
\def\grls{\;\lower4pt\hbox{${\buildrel\displaystyle >\over <}$}\;}

\title[Unstable Gravity Modes in Stellar Collapses]
{Adiabatic Perturbations in Homologous Conventional Polytropic
Core Collapses of a Spherical Star}
\author[Y. Cao and Y.-Q. Lou]{Yi Cao$^1$\thanks{y-cao04@mails.tsinghua.edu.cn}
and Yu-Qing Lou$^{1,\ 2,\ 3}$\thanks{louyq@mail.tsinghua.edu.cn; \ \ lou@oddjob.uchicago.edu }\\
    $^1$ Department of Physics and Tsinghua Centre for Astrophysics (THCA),
    Tsinghua University, Beijing 100084, China\\
    $^2$ Department of Astronomy and Astrophysics, the University of
    Chicago, 5640 S. Ellis Ave, Chicago, IL 60637, USA\\
    $^3$ National Astronomical Observatories, Chinese Academy of
    Sciences, A20, Datun Road, Beijing 100021, China
}
\date{Accepted 2009 November 30. Received 2009 November 29;
in original form 2009 September 22}

\begin{document}
\maketitle
\begin{abstract}
We perform a non-radial adiabatic perturbation analysis on
homologous conventional polytropic
stellar core collapses.
The core collapse features a polytropic exponent $\Gamma=4/3$
relativistic gas under self-gravity of spherical symmetry while
three-dimensional perturbations involve an adiabatic exponent
$\gamma$ with $\gamma\neq\Gamma$ such that the Brunt-V$\ddot{\rm
a}$is$\ddot{\rm a}$l$\ddot{\rm a}$ buoyancy frequency ${\cal N}$
does not vanish. With proper boundary conditions, we derive
eigenvalues and eigenfunctions for different modes of
oscillations. In reference to stellar oscillations and earlier
results, we examine behaviours of different modes and the
criterion for instabilities.
The acoustic p$-$modes and surface f$-$modes remain stable. For
$\gamma<\Gamma$, convective instabilities appear as unstable
internal gravity g$^{-}-$modes. For $\gamma>\Gamma$,
sufficiently low-order internal gravity g$^{+}-$modes are stable,
whereas sufficiently high-order g$^{+}-$modes, which would have
been stable in a static star, become unstable during self-similar
core collapses. For supernova explosions, physical consequences of
such inevitable g$-$mode instabilities are speculated.
\end{abstract}


\section{Introduction}

Stability properties of core collapses in massive progenitor stars
before supernova (SN) explosions (Goldreich \& Weber 1980 -- GW
hereafter; Goldreich et al.
1996; Lai 2000; Lai \& Goldreich 2000; Blondin et al.
2003; Murphy et al.
2004; Burrows et al. 2006, 2007; Lou \& Cao 2008; Cao \& Lou 2009)
have come to focus after three decades, because of the realization
after numerous unsuccessful one-dimensional SN simulations that
the breakdown of spherical symmetry inevitably occurs and plays a
key role in SN explosions (e.g. Burrows 2000). Numerical
simulations (e.g. Bruenn 1989a, b)
indicate that pre-SN stellar core collapses may be approximately
described by a homologous process, first analyzed by GW for a
conventional polytropic equation of state (EoS)
$P=\kappa\rho^{\Gamma}$ where $P$ and $\rho$ are the pressure and
density and both $\kappa$ and $\Gamma=4/3$ are constant. For
perturbations obeying identical EoS of the background core
collapse, GW explored linear stability properties of such stellar
collapses and concluded that these collapses are stable. Yahil
(1983) extended homologous collapses to polytropic exponent
$\Gamma<4/3$, noting the presence of an outer supersonic envelope
besides the inner core collapse. Lai (2000) performed perturbation
analysis to these extended solutions and claimed that
perturbations are stable for $\Gamma>1.09$.

Lou \& Cao (2008) substantially extended the $\Gamma=4/3$
self-similar solutions, including those of GW, using a general
polytropic EoS with a temporally and radially variable $\kappa$
being conserved along streamlines.
We obtained a broad family of homologous core collapses allowing
$\kappa$ as an arbitrary function of the independent self-similar
variable.

The specific entropy of an ideal or perfect gas is
\begin{eqnarray}
s=\frac{k_B}{(\gamma-1)}\ln\left(\frac{P}{\rho^\gamma}\right)+{\rm
constant}\ ,
\end{eqnarray}
where $k_B$ is the Boltzmann constant and $\gamma$ is the
adiabatic exponent with $\gamma\neq\Gamma$.
Here $\Gamma$ determines the structure and evolution of core
collapses while $\gamma$ controls specific entropy perturbation
properties. The earlier isentropic assumption requires
$\gamma=\Gamma$ (e.g. GW).
Numerical simulations (e.g. Bethe et al. 1979; Bruenn 1985, 1989a,
1989b; Woosley et al. 1993, 2002)
for the structure of massive stars and SN explosions support
variable radial distributions of specific entropy. Thus
the model of Lou \& Cao (2008) allows an arbitrary radial profile
of specific entropy. Using this general polytropic model, we
conducted a non-radial adiabatic perturbation analysis and
classified different perturbation modes parallel to stellar
oscillations (see Unno et al. 1979 and Cao \& Lou 2009)
and concluded that in addition to internal gravity g$^{-}-$modes
for convective instabilities, sufficiently high-order
g$^{+}-$modes also become unstable.

Cao \& Lou (2009) demonstrated that the criterion of using the
sign of the Brunt-V$\ddot{\rm a}$is$\ddot{\rm a}$l$\ddot{\rm a}$
buoyancy frequency squared ${\cal N}^2$ (see definition
\ref{defbv}) for the existence of g$-$modes remains valid in
self-similar dynamic core collapses. In a conventional polytropic
process, isentropic perturbations with $\gamma=\Gamma$ make all
g$-$modes disappear, i.e. ${\cal N}^2=0$.
Thus previous perturbation analyses (GW; Lai 2000; Lai \&
Goldreich 2000) considered only the stability of acoustic modes.
In addition to p$-$modes and f$-$modes, Cao \& Lou (2009) found
g$-$mode instabilities in stellar core collapses and speculated
possible consequences for neutron star kicks etc.

From the existence criterion of ${\cal N}^2\neq 0$ for g$-$modes,
we realize that in a conventional polytropic core collapse,
non-radial adiabatic perturbations with $\gamma\neq\Gamma$ should
allow g$-$modes.
Such nonisentropic assumption has been invoked decades ago (e.g.
Ledoux 1965; Unno et al. 1979) and applied to helioseismology and
white dwarf oscillations (e.g. Shibahashi et al. 1988).
For star formation, McKee \& Holliman (1999) also studied
nonisentropic radial perturbations in molecular clouds. They used
a thermal free energy to discuss cloud stability and gave critical
cloud masses for locally and globally adiabatic models. These
early results motivate us to perform such a perturbation analysis
for stellar core collapses and study the existence and stability
of g$-$modes in addition to other modes. The main thrust of this
Letter is to show such g$-$mode instabilities and speculate
consequences for stellar core collapses and SN explosions.

\section{Perturbations in core collapses}

The governing equations include conservations of mass and
momentum, Poisson equation for self-gravity and EoS.
The gas is ideal and
the background is a $\Gamma=4/3$ conventional polytrope while the
adiabatic perturbation with $\gamma\neq\Gamma$ conserves the
specific entropy along streamlines.
The background collapse is spherically symmetric
and as in GW, we introduce a dimensionless independent variable
${\bf x}={\bf r}/a(t)$ where $a(t)=\rho_c^{-1/3}[\kappa/(\pi
G)]^{1/2}$ is the Jeans length with a central mass density
$\rho_c(t)$ and $G$ is the gravitational constant. Physical
variables are cast into the following forms where the first term
of each variable is the dynamic background and the last term is
the first-order perturbation:
\begin{eqnarray}
{\bf u}=\dot{a}(t){\bf x}+\frac{a(t)}{t_{ff}}{\bf v}_1
(x,\ \theta,\ \varphi)\tau(t)\ ,\label{p1}\\
\rho=\left(\frac{\kappa}{\pi G}\right)^{3/2}a(t)^{-3}
f^3(x)[1+f_1(x,\ \theta,\ \varphi)\tau(t)]\ ,\label{p2}\\
P=\frac{\kappa^3}{(\pi G)^2}a(t)^{-4}f^4(x)[1
+\beta_1(x,\ \theta,\ \varphi)\tau(t)]\ ,\label{p3}\\
\Phi=\frac{4}{3}\left(\frac{\kappa^3}{\pi
G}\right)^{1/2}a(t)^{-1}[\psi(x)+\psi_1(x,\ \theta,\
\varphi)\tau(t)]\ .\label{p4}
\end{eqnarray}
Here the time-dependent temporal factor $\tau(t)$ is
\begin{equation}
\tau(t)=\exp\left(p\int^{t}t_{ff}^{-1}dt'\right)\ ,
\end{equation}
where the free-fall timescale $t_{ff}$ is defined by
\begin{equation}
t_{ff}=\left[\frac{4}{3}\pi G\rho_c(t)\right]^{-1/2}\ 
\end{equation}
and $p$ is a dimensionless `frequency'.
We do not use stream function for velocity perturbations to allow
for vorticities. The dynamic collapse background is that of GW.
Defining a dimensionless collapse parameter $\lambda$ (Yahil 1983)
as
\begin{equation}
\frac{1}{\lambda}\equiv\frac{8\pi G\rho_0
r^2}{3u_0^2}\bigg|_{r=0}=\left(\frac{u_{ff}}
{u_0}\right)^2\bigg|_{r=0}\
,
\end{equation}
where the subscript $0$ indicates background variables and
$u_{ff}$ is the free-fall velocity, we obtain the background
density profile determined by the equation for $f(x)$, namely
\begin{equation}\label{core}
\frac{1}{x^2}\frac{d}{dx}\left[x^2\frac{df(x)}{dx}\right]+f^3=\lambda\
,
\end{equation}
with boundary conditions $f(0)=1$ and $f'(0)=0$. Other background
variables are then readily determined. Solutions of eq
(\ref{core}) are sensible for core collapses of massive stars when
density vanishes at a moving boundary $x_b$.
For this purpose, the variable $\dot{a}$ should be negative and
the time reversal operation must be taken in the governing
equations. We assume adiabatic perturbations with the exponent
$\gamma$ different from $\Gamma=4/3$.
In our model, we define ${\bf v}_1=p{\bf w}_1$ and
$m=p[p+(\lambda/2)^{1/2}]$ for convenient analysis. After the
linearization of hydrodynamic equations including conservations of
mass and momentum, Poisson equation for self-gravity and the
adiabatic EoS, the angular variation factors of ${\bf w}_1$,
$\beta_1$, $f_1$ and $\psi_1$ can be separated out involving the
spherical harmonics $Y_{l\mathfrak{m}}(\theta,\ \phi)$.
We further let ${\bf w}_1$ take the consistent form of
\begin{equation}
 {\bf w}_1={\bf
e}_rw_r(x,t)Y_{l\mathfrak{m}}+w_t(x,t)\hat{\nabla}_\bot
Y_{l\mathfrak{m}}
+\hat{\nabla}_\bot\times(w_{rot}Y_{l\mathfrak{m}}{\bf e}_r)\ ,
\end{equation}
where the modified transverse gradient operator is
\begin{equation}
\hat{\nabla}_\bot\equiv{\bf
e}_\theta\frac{\partial}{\partial\theta}+{\bf
e}_\phi\frac{1}{\sin\theta}\frac{\partial}{\partial\phi}\ .
\end{equation}
As $w_{rot}$
decouples from other perturbation equations, we suppress it in the
following analysis. Eliminating $\beta_1$ and $f_1$, we obtain a
set of ordinary differential equations (ODEs):
\begin{eqnarray}
\frac{f}{x^2}\frac{d}{dx}\left(x^2w_r\right) -\frac{l(l+1)f}{x}w_t
+\frac{4}{\gamma}w_r\frac{df}{dx}\qquad\nonumber\\
-\frac{4}{3\gamma}(mxw_t+\psi_1)=0\ ,\label{eq1}\\
\frac{1}{x^2}\frac{d}{dx}\left(x^2\frac{d\psi_1}{dx}\right)
-\frac{l(l+1)}{x^2}\psi_1=3f^2\left(\frac{4}
{\gamma}-3\right)w_r\frac{df}{dx}\nonumber\\
-\frac{4}{\gamma}f^2(mxw_t+\psi_1)\ ,\label{eq2}\\
mw_r-m\frac{d}{dx}\left(xw_t\right)= \frac{(3\gamma-4)}{\gamma
f}\frac{df}{dx}(mxw_t+\psi_1)\qquad\nonumber\\
+\frac{3(4-3\gamma)}{\gamma
f}\left(\frac{df}{dx}\right)^2w_r\ ,\label{eq3}
\end{eqnarray}
where the same notation is used for the $x-$dependent factor of
$\psi_1$. ODEs $(\ref{eq1})-(\ref{eq3})$ for non-radial
perturbations reduce to those of GW by setting $\gamma=\Gamma=4/3$
and introducing a stream function for velocity perturbations.

Regular boundary conditions are imposed, namely
\begin{eqnarray*}
\left\{
\begin{array}{c}
\ \ \ \psi_1\propto x^l\ ,\qquad\quad w_r=lw_t\ ,
\qquad\qquad{\rm
for\ } x\rightarrow 0^{+}\qquad\qquad\\
\\
\psi_1\propto x^{-(l+1)}\ ,\quad 3w_rdf/dx-mxw_t=\psi_1\ ,
\quad {\rm for\ }x=x_b\
\end{array}
\right.\ .
\end{eqnarray*}
The last of the outer boundary conditions requires a zero
Lagrangian pressure. With these boundary conditions, we can prove
the orthogonality of eigenfunctions, i.e. $\int f^3{\bf
w}^{(i)}\cdot{\bf w}^{(j)}d^3r=0$ where superscripts $^{(i)}$ and
$^{(j)}$ indicate eigenfunctions of different eigenvalues.
Meanwhile, the variational principle (Chandrasekhar 1964) can be
formulated for this eigenvalue problem.

To solve this eigenvalue problem, we use the following numerical
procedure. After solving the background physical variables by
using the fourth-order explicit Runge-Kutta scheme, we discretize
ODEs (\ref{eq1})$-$(\ref{eq3}) with a proper mesh. Then the
inverse iteration method (e.g. Wilkinson 1965) is applied to
determine eigenvalues and eigenfunctions. To enhance the
efficiency of the inverse iteration method, a relaxation scheme is
also implemented.

\section{Results of Our Model Analysis}

We first test our numerical code to solve $\gamma=4/3$ isentropic
cases and confirm the results of Cao \& Lou (2009) to guarantee
the correctness and reliability of the code. This check also
confirms the errors of p$-$mode eigenvalues in GW.

\begin{figure}
\includegraphics[width=0.5\textwidth]{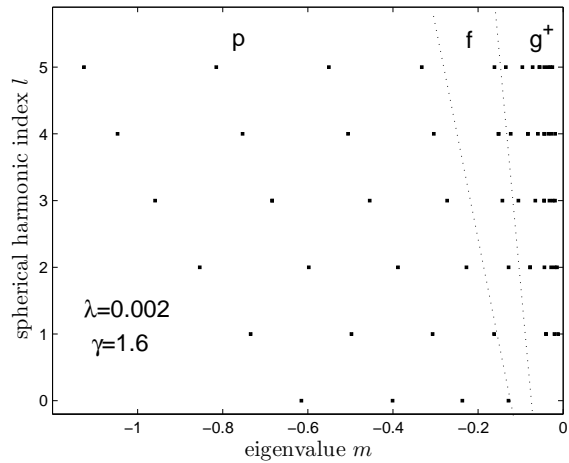}
\caption{Sample spectra of eigenvalues $m$ versus spherical
harmonic degree $l$ are shown with the parameter $\lambda=0.002$
and the adiabatic exponent of perturbations $\gamma=1.6$. All
eigenvalues lie in the $m<0$ regime. Two dotted lines divide the
region into three distinct zones for eigenvalues of different
modes. There is a sign difference between the definitions of the
eigenvalue $m$ here and in stellar oscillations (see figures of
Cox 1976 for a comparison).}\label{figure1}
\end{figure}

\begin{figure}
\includegraphics[width=0.5\textwidth]{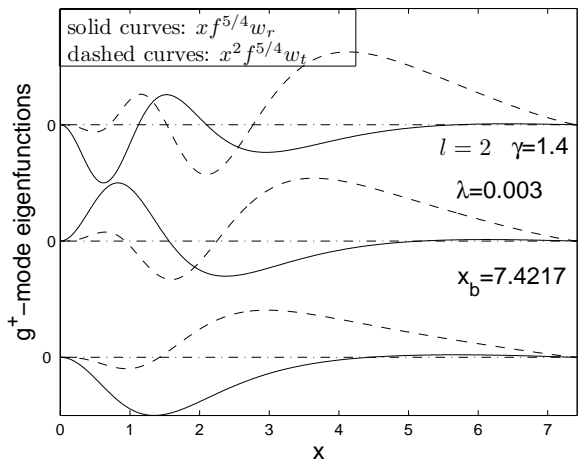}
\caption{Eigenfunctions of g$^{+}-$modes with $1,2,3$ nodes from
bottom to top panels with eigenvalues $m=-0.0249,\ -0.0135,$ $\
-0.0085$, respectively, are presented here. The solid and dashed
curves represent $xf^{5/4}w_r$ and $x^2f^{5/4}w_t$, respectively.
Factors $xf^{5/4}$ and $x^2f^{5/4}$ are multiplied for compact
presentations.}\label{figure3}
\end{figure}

The spectra of eigenvalues versus the mode degree $l$ contain
characteristic features of oscillations. A typical spectrum of
$\gamma>4/3$ modes is shown in Fig. \ref{figure1}. In these cases,
all eigenvalues $m$ are negative. We use two dashed lines to
separate three distinct classes of modes: p$-$modes for $l\ge 0$,
a unique branch of f$-$modes for $l\ge 2$ and g$^{+}-$modes for
$l\ge 1$ (see Cowling 1941 for mode classifications of stellar
oscillations and Unno et al. 1979 for properties of different
modes.).
The p$-$modes are pressure-driven and
gravity-modified acoustic oscillations. The f-modes are
essentially trapped surface Lamb waves (Lamb 1932; Unno et al.
1979; Lou 1990, 1991) and decay exponentially inwards. The
g$-$modes are driven by buoyancy and are trapped deep inside the
core. The g$^{+}-$modes are one subclass of g$-$modes with
$\gamma>4/3$.
For a given $l$, the absolute value of eigenvalue $|m|$ decreases
as the radial order increases. In that limit, eigenvalue $m$
approaches $0$. There is no $l=0$ g$^{+}-$mode. The other subclass
g$^{-}-$modes exist for $\gamma<4/3$ with similar characteristics
of g$^{+}-$modes, except that the eigenvalues $m$ are positive and
decrease towards zero as the radial order increases. By numerical
explorations, we find similar behaviours of each perturbation mode
in the core collapsing background as those in stellar
oscillations. As p$-$modes have been studied by GW for
$\gamma=\Gamma=4/3$, we emphatically show g$^{+}-$mode
eigenfunctions and their eigenvalues in Fig. \ref{figure3}.


Between p$-$modes and g$^{+}-$modes exists the unique branch of
f$-$modes for $l\ge 2$. Eigenvalues of f$-$modes separate the p$-$
and g$-$modes. Such f$-$modes are characterized by eigenfunctions
of density and radial velocity perturbations without nodes. They
are acoustic in nature and relate to the surface Lamb waves (Lamb
1932;
Lou 1990, 1991).


In stellar oscillations, the existence criterion for both types of
g$-$modes depends on the square of the Brunt-V$\ddot{\rm
a}$is$\ddot{\rm a}$l$\ddot{\rm a}$ buoyancy frequency ${\cal N}^2$
defined by
\begin{eqnarray}
{\cal N}^2=-\frac{1}{\rho}\frac{dP}{dr}\left(\frac{d\ln\rho}{dr}
-\frac{1}{\gamma}\frac{d\ln P}{dr}\right)\ ,\label{defbv}
\end{eqnarray}
where $P(r)$ and $\rho(r)$ are for a hydrostatic equilibrium, and
$\gamma$ is the adiabatic exponent of perturbation.
Since our
numerical explorations reveal that if an eigenvalue $m$ with its
eigenfunction can be obtained for $\lambda=0$ for the hydrostatic
limit, its counterpart can also be determined for $\lambda>0$ in a
continuous manner. Thus, the existence criterion for g$-$modes in
stellar oscillations can be applied in our analysis by using
partial derivatives instead of derivatives with respect to the
radius. For $\gamma<4/3$ (${\cal N}^2<0$), g$^{-}-$modes appear,
whereas for $\gamma>4/3$ (${\cal N}^2>0$), g$^{+}-$modes manifest.
For $\gamma=\Gamma=4/3$ (${\cal N}^2=0$), both g$-$modes are
suppressed.


We have readily computed the p$-$mode eigenfunctions with
$\gamma\neq\Gamma=4/3$ and they appear qualitatively similar to
those shown in GW and in Cao \& Lou (2009).

\begin{figure}
\includegraphics[width=0.5\textwidth]{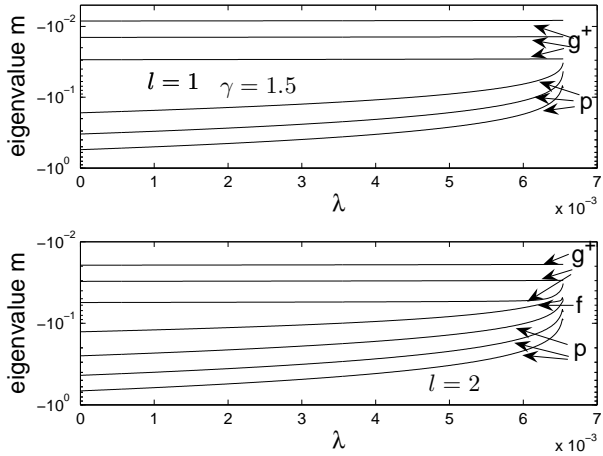}
\caption{Displayed are eigenvalue $m$ variations of the three
lowest acoustic p$-$modes and internal gravity g$^{+}-$modes of
$l=1,\ 2$ and of surface f$-$modes for $l=2$ with the collapse
parameter $\lambda$ characterizing different self-similar dynamic
backgrounds. The adiabatic exponent of perturbations is
$\gamma=1.5$.}\label{figure4}
\end{figure}

\begin{figure}
\includegraphics[width=0.5\textwidth]{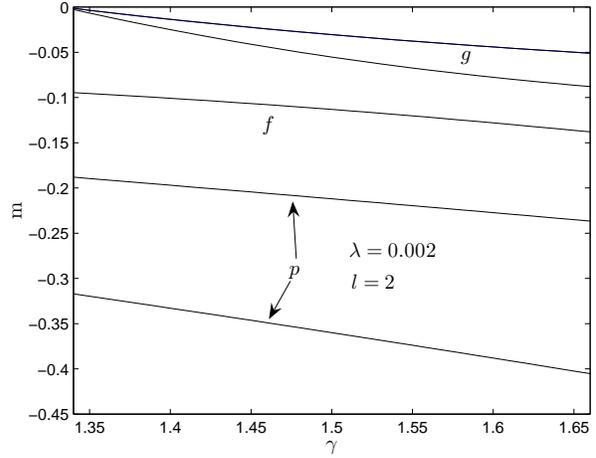}
\caption{Displayed are eigenvalue $m$ variations of the two lowest
acoustic p$-$modes and internal gravity g$^{+}-$modes and of
surface f$-$modes for $l=2$ with the adiabatic exponent $\gamma$
varying from $1.34$ to $1.66$. The collapse parameter is
$\lambda=0.002$ for the conventional polytropic stellar core
collapse of $\Gamma=4/3$.}\label{figure5}
\end{figure}

We demonstrate mathematically that a g$^{+}-$mode does not
change to a g$^{-}-$mode or {\it vice versa}
as $\lambda$ increases from $0$ to $\lambda_M$, i.e. for
$\gamma\neq 4/3$,
there is no $m=0$ eigenvalue for all $0<\lambda<\lambda_M$. By
setting $m=0$ in ODEs (\ref{eq1})$-$(\ref{eq3}) and after
straightforward manipulations, we derive
\begin{eqnarray}
\int_V\left[\left|\nabla\left(\psi_1Y_{l\mathfrak{m}}
\right)\right|^2 +3f^2\psi_1^2Y_{l\mathfrak{m}}^2\right]d^3r=0\ ,
\end{eqnarray}
which requires $\psi_1=0$. Thus no non-trivial eigenfunctions can
be found. Consequently, the eigenvalue curve $m(\lambda)$ for each
perturbation mode does not intersect the $m=0$ line. We show the
variation trends of eigenvalues $m$ as $\lambda$ increases from
$0$ to $\lambda_M$ in Fig. \ref{figure4} for typical eigenvalues
$m$ of the three lowest p$-$modes and g$-$modes for $l=1,\ 2$ and
of f$-$mode
for $l=2$ as functions of $\lambda$. Fig. \ref{figure4} shows that
eigenvalues $m$ larger than $25\lambda_M/8$ approach this limiting
value as $\lambda\rightarrow\lambda_M$ (see GW). Fig.
\ref{figure5} illustrates the variation of eigenvalues $m$ for the
first two lowest orders of p$-$modes and g$^{+}-$modes as well as
that of f$-$modes all for $\lambda=0.002$ and $l=2$ as the
adiabatic exponent $\gamma$ varies from $1.34$ to $1.66$.

A key question for perturbation analysis is the stability of a
core collapse.
The inequality ${\cal N}^2<0$ for stellar perturbations is the
Schwarzschild criterion for convection. In our dynamic core
collapse background, we emphasize that this inequality is only
sufficient but not necessary for instability. The temporal factor
$\tau(t)$ for $\lambda> 0$ bears a power law $t^{\xi}$ with
$\xi=-1/6\pm [1/36+2m/(9\lambda)]^{1/2}$.
For a time reversal operation, GW
showed that the compression of a collapse is responsible for an
amplification of $t^{-1/6}$. Thus the determinant under the square
root in $\xi$ decides the mode stability. For $m>-\lambda/8$ so
that $\xi$ has two real roots, the mode is unstable for one of the
two roots. For $m<-\lambda/8$ so that $\xi$ has a pair of complex
conjugate roots, a perturbation oscillates stably in addition to
the compression amplification of core collapse.
By this new criterion for instability, we find that all
g$^{-}-$modes and sufficiently high-order g$^{+}-$modes are
unstable.
It can be shown by a local analysis that the
definition of ${\cal N}$ remains valid for dynamical core
collapses, such that g$^{-}-$modes lead to convective
instabilities. It also announces that g$^{+}-$mode instabilities
are uniquely associated with self-similar dynamic core collapse.

We compare our results with earlier analyses (GW; Lai 2000; Cao \&
Lou 2009).
First, in contrast to oscillations in a static
star, we examine non-radial adiabatic perturbations in a
self-similar conventional polytropic core collapse. The adiabatic
exponent $\gamma$ differs from the background $\Gamma=4/3$,
whereas earlier stability analyses of the dynamic background are
restricted to $\gamma=\Gamma=4/3$. Secondly, for a conventional
polytropic core collapse,
we classify different adiabatic perturbation modes, including
p$-$modes, g$-$modes and f$-$modes.
GW and Lai (2000) studied only p$-$mode perturbations as their
isentropic EoS makes g$-$modes vanish. Cao \& Lou (2009) obtained
g$-$modes by a general polytropic EoS with a variable specific
entropy distribution, even though the perturbation EoS remains the
same as the background EoS. In comparison, our model here
describes non-radial adiabatic perturbations with
$\gamma\neq\Gamma=4/3$. This leads to ${\cal N}\neq 0$ so that
g$-$modes may exist. We indeed confirm this by numerical
explorations.
For various efficiencies of heat transport and radiative losses,
the case of $\gamma=\Gamma$ should be a special and rare
situation. Therefore g$-$modes should exist in stellar core
collapses in general.

For core-collapse SN explosions, such g$-$mode instabilities
should bear physical consequences. These instabilities complement
those
revealed by Cao \& Lou (2009). They occur during the pre-SN core
collapse stage and should also influence the formation of
proto-neutron stars and subsequent emergence and evolution of
rebound shocks. At least, such instabilities lead to early
breakdown of spherical symmetry before the emergence of rebound
shocks (e.g. Lou \& Wang 2006, 2007; Wang \& Lou 2008; Hu \& Lou
2009). So far, most proposed instabilities occur either in a
massive progenitor star prior to the core collapse (e.g. Goldreich
et al. 1996) or after the core rebound (e.g. Blondin et al. 2003;
Burrows et al. 2006, 2007). Such instabilities may affect the
formation, motion and evolution of a proto-neutron star or a
pulsar as speculated by Cao \& Lou (2009). Moreover, our
perturbation analysis serves to link perturbations in a massive
progenitor star and those after the core rebound. For example, the
$`\epsilon$-mechanism' (e.g. Goldreich et al. 1996) may provide
seed perturbations during a core collapse while such perturbations
during a core collapse may stimulate those after the core rebound.

We emphasize that while p$-$modes, f$-$modes and low-order
g$^{+}-$modes are stable, if radiative losses and diffusive
processes are involved instead of the adiabatic approximation,
these modes might become overstable.

\section{Summary and conclusions}

In this Letter, we examine 3D
adiabatic perturbations in a self-similar conventional polytropic
collapsing core with $\gamma\neq\Gamma=4/3$.
The gas is non-isentropic in the sense that the adiabatic index
$\gamma$ of perturbations differs from that of the background
$\Gamma=4/3$. As $\gamma$ determines the specific entropy
perturbation conserved along streamlines, the non-isentropic
process actually involves
a nonzero buoyancy frequency ${\cal N}$ giving rise to
g$-$mode perturbations.
In comparison, perturbation analysis of
Cao \& Lou (2009) emphasizes a general polytropic core collapse
with a variable specific entropy (Lou \& Cao 2008) and 3D
adiabatic perturbations of $\gamma=\Gamma=4/3$.

By imposing proper boundary conditions, we solve the perturbation
eigenvalue problem and derive distinct modes: acoustic p$-$modes,
surface f$-$modes, internal gravity g$^{-}-$ and g$^{+}-$modes
which are classified by their series of eigenvalues $m$ and
eigenfunctions in reference to stellar oscillations. In parallel,
g$-$modes involve two types, viz. g$^{+}-$ and g$^{-}-$modes;
their existence depends on the square of the Brunt-V$\ddot{\rm
a}$is$\ddot{\rm a}$l$\ddot{\rm a}$ buoyancy frequency ${\cal
N}^2$:
the former requires ${\cal N}^2>0$ while the latter needs ${\cal
N}^2<0$. For adiabatic perturbations, g$^{+}-$modes correspond to
$\gamma>4/3$ while g$^{-}-$modes occur for $\gamma<4/3$.

Stability properties of these modes in stellar core collapses are
examined. The instability criterion shifts from $m>0$ in a static
Lane-Emden polytropic sphere of $\lambda=0$ to $m>-\lambda/8$ for
$0<\lambda<\lambda_M$ core collapses. Consequently, p$-$modes,
f$-$modes and sufficiently low-order g$^{+}-$modes oscillate
stably. The g$^{-}-$modes are unstable leading to convections.
Sufficiently high-order g$^{+}-$modes which would have been stable
in a static polytrope now become unstable for
$0<\lambda<\lambda_M$ GW dynamic core collapse. The specific
radial order that g$^{+}-$modes become unstable depends on the
parameter pair of $\lambda$ and $\gamma$.

We speculate that such inevitable g$-$mode instabilities may offer
valuable clues to SN simulations and that the formation, motion
and evolution of proto-neutron stars can be nontrivially
influenced (Cao \& Lou 2009). These perturbations again lead to
instabilities before the core rebound and they serve as seeds of
later fluctuations. Their excitations can come from oscillations
of the progenitor star before the onset of core collapse.
According to our numerical exploration and typical stellar
parameters with $\gamma>\Gamma=4/3$, unstable g$^{+}-$modes appear
with fairly high radial orders. This implies that the central mass
blob can be quite small and a SN might even break the core into
pieces without forming a NS.


\section*{Acknowledgements}

This research was supported in part by Tsinghua Centre for
Astrophysics, by the National Natural Science Foundation of China
grants 10373009, 10533020 and J0630317
at Tsinghua University, and by the SRFDP 20050003088 and
200800030071, the Yangtze Endowment and the National Undergraduate
Innovation Training Project from the Ministry of Education at
Tsinghua University.

\end{document}